\documentclass[12pt]{article}
\usepackage{epsfig,amsfonts,amssymb}
\usepackage{hyperref}
\usepackage{cite}
\input epsf.sty
\usepackage{cite}

\newcommand{\ben}{\begin{eqnarray}\displaystyle}
\newcommand{\een}{\end{eqnarray}}
\newcommand{\be}{\begin{equation}}
\newcommand{\ee}{\end{equation}}
\newcommand{\lb}{\left (}
\newcommand{\rb}{\right )}
\newcommand{\ltb}{\left [}
\newcommand{\rtb}{\right ]}

\newcommand{\ra}{\rightarrow}

\begin{document}

\baselineskip 24pt

\begin{center}
{\Large \bf Near-Horizon Analysis of $\eta/s $}

\end{center}

\vskip .6cm
\medskip

\vspace*{4.0ex}

\baselineskip=18pt

\centerline{\large \rm   Nabamita Banerjee$^{(1)}$, Suvankar Dutta$^{(2)}$}

\vspace*{4.0ex}

\centerline{\large \it $^{(1)}$ Institute for Theoretical Physics, Utrecht University}

\centerline{\large \it Leuvenlaan 4, 3584 CE, Utrecht, The Netherlands}

\vspace*{4.0ex}

\centerline{\large \it $^{(2)}$ Dept. of Physics, Swansea University}

\centerline{\large \it  Swansea, UK}

\vspace*{1.0ex}
\centerline{E-mail: N.Banerjee at uu dot nl, pysd at swan dot ac dot uk}

\vspace*{5.0ex}

\centerline{\bf Abstract} \bigskip

It is now well understood that the coefficient of shear viscosity of boundary fluid can be obtained from the horizon values of the effective coupling of transverse graviton in bulk spacetime. In this paper we observe that to find the shear viscosity coefficient it is sufficient to know only the near horizon geometry of the black hole spacetime. One does not need to know the full analytic solution. We consider several examples including non-trivial matter (dilaton, gauge fields) coupled to gravity in presence of higher derivative terms and calculate shear viscosity for both extremal and non-extremal black holes only studying the near horizon geometry. In particular, we consider higher derivative corrections to extremal R-charged black holes and compute $\eta/s$ in presence of three independent charges. We also consider asymptotically Lifshitz spacetime whose dual black hole geometry can not be found analytically. We study the near horizon behaviour of these black holes and find $\eta/s$ for its dual plasma at Lifshitz fixed point.

\vfill \eject

\baselineskip=18pt

\tableofcontents

\section{Introduction and Summary}\label{intro}

Black holes have strange thermodynamic behaviour. Unlike other thermodynamic objects its entropy is proportional 
to {\it area} (of the event horizon) instead of its volume. 
\be
S= {A_{d-1} \over 4 G_{d+1}},
\ee
where $A_{d-1}$ is $d-1$ dimensional horizon (spatial)area and $G_{d+1}$ is Newton's constant in $d+1$ dimension. 
Therefore all the microscopic details of a black hole are encoded in one less dimension. 
It was first observed by Bekenstein for two derivative gravity. Later Wald has prescribed a generalized covariant formula for non-extremal black hole entropy in presence any higher derivative terms. Though the area law does not hold in higher derivative gravity but to calculate entropy using Wald's formula we only require the horizon values of metric and other fields forming the black holes. In the extremal limit one has to be careful using Wald's formula since the bifurcate killing horizon does not exist in this case. In \cite{sen} Sen has discovered an elegant method to calculate the entropy for a wide class of extremal black holes. Again, only the near horizon values of the fields are required to calculate entropy.     

Black holes in asymptotically AdS space is more interesting. AdS/CFT predicts that the thermodynamic and hydrodynamic behaviour of strongly coupled boundary plasma are captured by black holes in one higher dimension. 
The entropy of a black hole in
AdS space is related to the thermodynamic entropy of the boundary
gauge theory at a finite temperature (which is the same as the
Hawking temperature of the black hole). Thus there are two
apparently different holographic descriptions of the entropy. One in
terms of the horizon and the other in terms of the boundary
gauge theory \cite{dg}. In a rough sense, the latter is the UV description from
the microscopic (gauge theory) point of view, while the former is the
IR description. One might therefore expect some kind of RG flow to relate the
two. 

Like entropy, the hydrodynamic quantities of boundary conformal plasma can be evaluated holographically. For example the retarded Green's function for tensor mode (for example $h_{xy}$) can be calculated using Kubo formula,
\be
G^R_{xx,xy} = P - i  \eta \omega + \eta \tau_{\pi} \omega^2 + {\kappa \over 2}[(p - 2)\omega^2 + q^2] 
+ {\cal O}(\omega^2),
\ee 
where, $p$ is spatial dimension and $p \geq 3$ and
\ben
P &:& pressure,\\
\eta &:& shear \ viscosity \ coefficient,\\
\tau_{\pi} &:& relaxation \ time \ for \ shear \ viscous \ tensor,
\een
and $\omega$ is frequency and $q$ is spatial momentum.

As we have seen that it is possible to find the the entropy of boundary field theory, only specifying the complete near horizon geometry of its dual black hole spacetime, one can also ask in the same spirit, whether 
different hydrodynamics coefficients of the boundary plasma can be found from the knowledge of bulk near-horizon geometry.
There are enough hints \cite{thorne,pw1,sonsh,liu} which seems to point to an affirmative answer to this question. In \cite{liu} it has been shown that in the low 
frequency limit ($\omega \ra 0$) the shear viscosity coefficient $\eta$ is related to transport 
coefficient of membrane fluid, which in turn is given by the effective coupling constant of graviton 
($h_{xy}$) evaluated on the horizon. In \cite{cai1,ns1} the prescription has been generalized for any higher 
derivative gravity\footnote{Recently in \cite{paulos} a Wald like formula for shear viscosity has been proposed.}. 
All these approaches indicate that there can be an "IR" description of transport coefficients $\eta$. However, it is not 
quite clear how other transport coefficients like $\tau_{\pi}$ etc. which appear in next order in $\omega$ can be 
calculated only in terms of horizon data. The derivation of \cite{liu,ns1} was strictly valid only in 
$\omega \ra 0$ limit.

Although, all the above references state that that one can possibly analyse the transport coefficient $\eta$ of the boundary plasma completely in terms of horizon data, but still the full bulk solutions were used in the actual computation. In this paper, we 
observe that only with the knowledge of near horizon geometry of a black hole spacetime one
can easily calculate the thermodynamic and hydrodynamic quantities, namely entropy and shear viscosity of boundary 
fluid and their ratio. One does not need to know the full analytic solutions of Einstein equations. This observation 
is helpful when bulk Lagrangian is very complicated. For example when gravity is coupled to various matter fields 
(gauge field, scalars etc.) in non-trivial way, it may not be always possible to find an analytic solution. 
But if a black hole solution exists then it is possible to find the near 
horizon geometry, i.e. how the metric and other fields behave in near horizon limit. The method is very simple. 
We do not need to solve any differential equation to find the near horizon geometry. First we write the field equations. 
Then take a suitable near horizon ansatz for different fields. In general different components of the metric goes like $g_{tt} \sim a_1 (r-r_h) + a_2 (r-r_h)^2+ \cdots, \ \ g_{rr}\sim {b_1 \over r-r_h}(1+ b_2 (r-r_h)+\cdots$)(for non-extremal case), scalar fields behaves as $\varphi \sim \varphi_h + \varphi_1 (r-r_h)+\cdots$ and gauge fields goes as $A_t = q(r-r_h)+\cdots$. Substituting these in the field equations we find the coefficients 
consistently order by order in $(r-r_h)$. Once we find the near-horizon structure of the spacetime we find entropy of the system using Wald's formula (for non extremal black hole) or entropy function formalism (for extremal black hole). In either case we only need to calculate the horizon values of some quantities. To calculate shear viscosity we adopt the method proposed in \cite{liu,cai1,ns1}. We write the effective action for transverse graviton in black hole near-horizon region and find the coupling constant. For non-extremal black hole usually (in presence of higher derivative terms) the effective coupling depends on radial coordinate and we need to take $r\ra r_h$ limit at the end to find the horizon value of the coupling constant. For extremal black hole it turns out to be constant.

In case of non-extremal black holes the near horizon geometry is not completely determined always.  
For example we consider dilaton field coupled to cosmological constant and higher derivative terms in sec. 
\ref{neutralisation} and dilaton field coupled to gauge field non trivially in sec. \ref{dilcharge}. We found 
that the horizon values of the dilaton field ($\varphi_h$) is not determined. Other coefficients of metric, gauge fields and dilaton depend on $\varphi_h$. 
Therefore the ratio $\eta/s$ depends on the horizon values of the dilaton field. Where as in extremal case the near-horizon geometry is completely fixed in terms of black hole charges and hence $\eta/s$ is also completely determined in  terms of the charges of black hole.

Our paper is organised as follows. In section \ref{nonex} we study the near-horizon geometry of different non-extremal black holes and find $\eta/s$ for dual conformal plasma in presence of global charges. In particular we studied $F^4$ terms in Gauss-Bonnet Gravity (sec. \ref{f4}), non-trivial dilaton coupled to Gauss-Bonnet term (sec. \ref{neutralisation}) and Maxwell term (sec. \ref{dilcharge}). We study near horizon geometries of extremal black holes in presence of different higher derivative terms and find $\eta/s$ of boundary field theory at zero temperature (sec. \ref{ex}). Finally in sec. \ref{lif} we consider near-horizon geometry of black holes in asymptotically Lifshitz spacetime and find $\eta/s$ for its dual plasma at Lifshitz fixed point.

\section{Non Extremal Black Holes}\label{nonex}

Now, we will consider theories with some matter coupled to metric. Here, we find $\eta/s $ for these theories by computing the horizon value of graviton's effective coupling. In \cite{ns1}, we proved that for generic higher-derivative term added to the Einstein-Hilbert action, one can actually write an effective action (which gives the same equation of motion as the original one) in canonical form. The canonical form of the action was required to conclude that the transport coefficients (in particular shear-viscosity coefficient) of the boundary plasma is related to the effective coupling of the transverse graviton. Hence, to justify that even in theories with arbitrary matter content, the effective coupling can still be related to the transport coefficient of the dual boundary plasma, we need to prove that the action would maintain its canonical form. This issue has been well studied in \cite{kovtun}. In presence of generic matter in the action, one can still show that the transverse graviton satisfies the equation of a massless scalar field. Therefore, the action evaluated on shell would always have the canonical form.

\subsection{$F^4$ Terms In Gauss-Bonnet Gravity}\label{f4}

As a first example we consider the following Lagrangian. 
\ben
{\cal I}&=&{1\over 16 \pi G_5} \int d^5 x \sqrt{-g} \Big{[} R + 12 + g \lb R^2 -4 R_{ij}R^{ij}+R_{ijkl}R^{ijkl} \rb \nonumber \\
&& -{1\over 4} F_{ij}F^{ij}  + c_1 (F_{ij}F^{ij})^2 + c_2 F_{ij}F^{jk}F_{kl}F^{li} \Big{]}\ .
\een

From the string theory point of view, the $F^4$ term is of order of $\alpha'$ as the Gauss-Bonnet term. 
In fact, in general one can also include other four derivative terms in the action\cite{myerscharge,cremonini} which arise in five dimensional gauge supergravity. However for simplicity we consider the above action. In \cite{cai1}, the authors have calculated shear viscosity to entropy density ratio using full bulk solution. Since the complete bulk solution is very complicated in presence of Gauss-Bonnet term it turns out to be difficult to express the ratio completely in terms of black hole charges. Here we study the near-horizon geometry of this black hole spacetime which is in fact very simple and compute shear viscosity, entropy and their ratio. We are able to express them in terms of black hole charges. Our expressions are exact in $g$.

We consider the following ansatz for metric and gauge field,
\ben
ds^2&=& -g_{tt}(r) dt^2 + {\sigma^2(r)\over g_{tt}(r)}dr^2 + r^2 d\vec{x}^2\\
A(r)&=&A_t(r) \ dt
\een
where,
\ben
g_{tt}(r)&=&a_1(r-r_h)+ a_2(r-r_h)^2 \cdots,  \nonumber \\
\sigma(r)&=&\sigma_0 + \sigma_1(r-r_h)+ \cdots,  \nonumber \\
A_t(r)&=& q \ \sigma_0(r-r_h)+ \cdots.
\een
The coefficient $\sigma_0$ can be set to one by rescaling the time and the parameter $q$ is related to the physical charge of the system.
Substituting this ansatz in field equations one can solve for $a_1$, $a_2$, $\sigma_1$ etc. order by order in ($r-r_h$). It turns out that only the coefficient $a_1$ appears in entropy, shear viscosity and their ratio. 
\be
a_1=-{r_h (6 c_1 q^4 + 3 c_2 q^4 + q^2 - 96)\over 24} \ .
\ee

Once we know the near horizon geometry one can use Wald's formula to find the entropy of the system. It is given by (s in entropy density),
\be
s={r_h^3 \over 4 G_5}\ .
\ee
Using the prescription given in \cite{liu,cai1,ns1} we calculate shear viscosity coefficient. It turns out to be,
\be
\eta = \frac{r_h^3}{16 \pi  G_5}\lb 1 - {2 g a_1\over r_h} \rb\ .
\ee
Therefore the ratio $\eta/s$ is given by,
\ben
{\eta \over s} &=&\frac{1}{4 \pi }\lb 1 - {2 g a_1\over r_h} \rb \nonumber \\
&=&\frac{1}{4 \pi }\lb 1 + { g}\ {6 c_1 q^4 + 3 c_2 q^4 + q^2 - 96\over 12}\rb\ .
\een

Temperature can also be calculated with near horizon data. It turns out to be,
\be
T={a_1 \over 4 \pi}
\ee
Therefore one can express $\eta/s$ in terms of black hole temperature,
\be
{\eta \over s} =\frac{1}{4 \pi }\lb 1- {8 \pi g \over r_h} {T}\rb\ .
\ee
This expression matches with \cite{cai1}. However here we are able to express $\eta/s$ in terms of charge parameter also. In the limit $T\ra 0$ (which is extremal limit), the ratio $\eta/s$ approaches its universal value.

\subsection{Non-trivial Dilaton}\label{neutralisation}

So far we have considered Maxwell-Einstein-Hilbert action in presence of four derivative terms. In this section we consider dilaton field which is non-minimally coupled to gravity and find the effects of non-trivial dilaton on $\eta/s$.
\be
{\cal I}= {1 \over 16 \pi G_{5}}\int d^5x \sqrt{- g} \ltb R -{1 \over 2}\partial_{\mu}\varphi \partial^{\mu}\varphi 
- 2 \Lambda \,\,e^{\tau \varphi}+{g}e^{- \gamma \varphi}\,\, {\cal L}_{(GB)}\rtb
\ee
with,
\be
{\cal L}_{(GB)}= R^2+ R^{\mu\nu\rho\sigma}R_{\mu\nu\rho\sigma} -4 R^{\mu\nu}R_{\mu\nu}
\ee
where, $\gamma$ and $\tau$ are constants and $\varphi$ is
the dilaton field. One can get this action by transforming the string frame action with
a non-trivial dilaton to the Einstein frame. Though from string theory point of view the values of $\tau$ and $\gamma$ are fixed but here we keep them as some arbitrary parameter.  In \cite{got} it has been shown that there exists an asymptotically AdS black hole solution for some suitable range of parameters. However here we examine the near horizon geometry of this black hole spacetime and calculate $\eta$, $s$ and their ratio. It turns out that the ratio (also $s$ and $\eta$) depends on the horizon values of the dilaton field. The horizon value of the dilaton field is not some arbitrary parameter of the theory it is fixed in terms of the charges of the black hole (in this case only mass). Our results are exact in $g$.

The near horizon ansatz for this system is given by,
\ben
ds^2&=& -B(r) e^{-2 \delta(r)} dt^2+{dr^2 \over B(r)}+ r^2 d\vec{x}^2,\nonumber \\
\delta(r)&=& \delta_0+\delta_1 (r-r_h)+ \delta_2 (r-r_h)^2+\cdots,\nonumber \\
B(r)&=& b_1 (r-r_h)+b_2(r-r_h)^2+\cdots,\nonumber \\
\varphi(r)&=& \varphi_h+ \alpha_1 (r-r_h)+\alpha_2(r-r_h)^2+\cdots\ .
\een
Here, $\varphi_h$ is the constant horizon value of the dilaton field. We substitute these ansatz in the equation of motion 
and get the following values of the coefficients (we present the expressions of those coefficients which appear in $\eta/s$).
\ben
\alpha_1 &=&-{8 e^{-(\gamma-\tau)}g \gamma \Lambda + 3 \tau \over r_h},\nonumber   \\
\delta_1&=& - {r_h \alpha_1^2 \over 6}, \nonumber \\
b_1&=&-{2 \over 3}e^{\tau \phi_h}r_h \Lambda .
\een

Once we know the near horizon geometry we calculate $s$ and $\eta$ as before and get,
\be
s= {r_h^3 \over 16 \pi G_5},
\ee
and
\be
\eta=
r_h^3\frac{e^{-2 \gamma  \varphi_h} \left(32 g^2 \gamma ^2
   \Lambda ^2 e^{2 \tau  \varphi_h}+4 g \Lambda  (3
   \gamma  \tau +1) e^{\varphi_h (\gamma +\tau )}+3
   e^{2 \gamma  \varphi_h}\right)}{48 \pi  G_5}\ .
\ee
Therefore the ration turns out to be,
\ben
{\eta\over s}&=&\frac{1}{4 \pi }+
\frac{8 g^2 \gamma ^2 \Lambda ^2 e^{2 \tau  \varphi_h-2
   \gamma  \varphi_h}}{3 \pi }+\frac{g \gamma  \Lambda 
   \tau  e^{(\tau - \gamma) 
   \varphi_h}}{\pi }+\frac{g \Lambda  e^{(\tau - \gamma) \varphi_h}}{3 \pi
   } \nonumber \\
&=& {1\over 4 \pi} \ltb 1-8 g e^{-(\gamma-\tau)\varphi_h} \lb 1+ 3 \gamma \tau - 48 \gamma^2 g 
e^{-(\gamma-\tau)\varphi_h} \rb \rtb\ , \ \ \ \Lambda = -6\ .
\een
The answer has been obtained in \cite{cai2}. 


\subsection{Charged Black Holes with Dilaton}\label{dilcharge}

The last non extremal black hole we consider is electrically charged black hole in presence of non-trivial dilaton field in presence generic four derivative terms. The action is given by \cite{gm},
\ben
{\cal I}={1 \over 16 \pi G_5}\int d^Dx \sqrt{-g}[ R &-& {1 \over D-2} \partial_{\mu}\phi 
\partial^{\mu}\varphi - V(\varphi) - {1 \over 4} e^{{- g_{0} \varphi}} F_{\mu\nu}F^{\mu\nu}\\
&-& {1 \over 2(D-2)!} e^{- g_{0} \varphi} F_{\mu_1\mu_2....\mu_{D-2}}F^{\mu_1\mu_2....
\mu_{D-2}}+ g {\cal L}_{(4)}]
\een
with,
\be
{\cal L}_{(4)} = \beta_1 R^2+ \beta_2 R^{\mu\nu\rho\sigma}R_{\mu\nu\rho\sigma}+\beta_3 R^{\mu\nu}R_{\mu\nu}
\ee
It has been shown in \cite{ptw} that when $V(\varphi)=|\Lambda|$ (positive cosmological constant) there exists not black hole solution. However there exists an asymptotically AdS spherical black hole solution when, $V(\varphi)=-|\Lambda|$ with a maximum single horizon. 
The asymptotic value of the dilaton field is fixed in terms of charges of the black hole. Therefore it is not an independent hair. In \cite{ptw} authors found a numerical black hole solution with two parameters namely position of the horizon $r_h$ and horizon value of the dilaton field $\varphi_h$. Here we find near-horizon geometry of the black hole spacetime and calculate $\eta/s$. The ratio depends on the horizon value of the dilaton. The results are valid up to order $g$.

We will work in five dimension and only with electrically charge gauge fields. We also treat the four derivative terms perturbatively. The leading (in $g\ra 0$ limit) near-horizon ansatz is
\ben
ds^2&=& -A(r) \sigma(r)^2 dt^2 +{dr^2 \over A(r)}+ r^2 dX^2, \nonumber \\
A(r)&=&\alpha_1 (r-r_h)+\alpha_2 (r-r_h)^2 +\cdots, \nonumber \\
\sigma(r)&=& \sigma_0+\sigma_1 (r-r_h)+\sigma_2 (r-r_h)^2 +\cdots ,\nonumber \\
\varphi(r)&=&\varphi_h+\varphi_1 (r-r_h)+\varphi_2 (r-r_h)^2 +\cdots ,\nonumber \\
A_t(r)&=&Q(r-r_h) + q_2(r-r_h)^2 + \cdots\ .
\een
The near-horizon behaviour of the fields as follows
\be
\alpha_1={r_h \over 24}\lb8 \Lambda - e^{-g_0 \varphi_h}Q^2\rb,\ \ \ \ \ \ \ \sigma_1 = {3 g_0^2 Q^4 
\sigma_0 \over 2 r_h (Q^2-8 e^{-g_0 \varphi_h}Q^2)} 
\ee
and
\be
 \alpha_2= \frac{e^{-g_0 \varphi_h} \left(\left(9 g_0^2+14\right) Q^4-128 Q^2 \Lambda  
e^{g_0 \varphi_h}+128 \Lambda ^2 e^{2 g_0
   \varphi_h}\right)}{96 \left(Q^2-8 \Lambda e^{g_0 \varphi_h}\right)}   .
\ee

The entropy and shear viscosity is given by,
\be
s={r_h^3 \over 4 G_5} \lb 1 + {e^{-g_0 \varphi_h}(-Q^2(\beta_1 +7\beta_2+2\beta_3) 
-8 e^{g_0 \varphi_h}(5\beta_1 -\beta_2+\beta_3)\Lambda )g \over 12}\rb + {\cal O}(g^2)
\ee
\be
\eta={r_h^3 \over 16 \pi G_5} \lb 1 + {e^{-g_0 \varphi_h}(-Q^2(\beta_1 +6\beta_2+2\beta_3) 
-8 e^{g_0 \varphi_h}(5\beta_1+\beta_3)\Lambda )g \over 12}\rb\ + {\cal O}(g^2).
\ee
Therefore the ratio is given by,
\be
{\eta\over s}={1 \over 4\pi}\ltb 1+ g \beta_2 {e^{-g_{0}\varphi_h} Q^2 - 8 \Lambda \over 12} \rtb\ + {\cal O}(g^2).
\ee
The black hole temperature can also be calculated from near-horizon geometry. 
\be
T\sim e^{-g_{0}\varphi_h} Q^2 - 8 \Lambda\ .
\ee
In $T\ra 0$ limit the the horizon value of the dilaton is also determined in terms of black hole charges and the ratio approaches its universal value.


\section{Extremal Black Holes}\label{ex}

We will first briefly discuss how charge black holes can arise in five
dimensions in the context of string theory.  A consistent truncation
of ${\cal N}=8$, $D=5$ gauged supergravity with $SO(6)$ Yang-Mills
gauge group, which can be obtained by $S^5$ reduction of type IIB
supergravity, gives rise to ${\cal N}=2$, $D=5$ gauge supergravity
with $U(1)^3$ gauge group.
The same theory can also be obtained by compactifying eleven
dimensional supergravity, low energy theory of M theory, on a
Calabi-Yau three folds. The bosonic part of the action of ${\cal
  N}=2$, $D=5$ gauged supergravity is given by, 
\ben\label{gsacn}
{\cal I} &=& {1 \over 16 \pi G_{5}}\int d^5x \sqrt{- g} \ltb R 
- {1 \over 4}X_I^2 F^I_{\mu\nu}F^{\mu\nu I}-{1 \over 2}X_I^{-2}
\partial_{\mu} X_I \partial^{\mu}X_I -V \rtb  \nonumber \\
&& + C.S \,\,\, terms
\een
where, $X^I$'s are three real scalar fields, subject to the constraint
$X^1 X^2 X^3 =1$. $F^I$'s, which are field strengths of three Abelian gauge
fields (I,J=1,2,3), and the scalar potential $V(X)$ is given by,
\ben
F_{\mu\nu}^I &=& 2 \partial_{[\mu}A^I_{\nu]},\,\,\,\,\ V= - 4 \gamma^2 \Sigma
_{I=1}^{3}X_I,\,\,\,\ X_1X_2X_3=1\ .
\een

Solving the equations of motion one gets a black hole solution with three $U(1)^3$ charges. Using this equations of motions one can calculate entropy and shear viscosity of boundary plasma and their ratio. It turns out that the ratio saturates the famous {\bf KSS} bound \cite{sonr}. 
We will consider higher derivative corrections to this action in next two subsections. Calculation of entropy and shear viscosity for non-extremal black hole for three independent non-zero charges seems to be very difficult. One can consider different limits, for example three equal charges (which is same as Maxwell-Einstein-Hilbert action) and calculate different thermodynamic and hydrodynamic coefficients. 
However we will concentrate only on the extremal limit of this black hole (for three independent charges). In the extremal limit the near horizon geometry is universal and given by \cite{morales}.  We calculate corrections to entropy for higher derivative terms using entropy function formalism. Evaluating the effective action for transverse graviton in this near horizon geometry we also compute higher derivative correction to shear viscosity coefficient of zero temperature boundary plasma in presence of global $U(1)^3$ charges.

\subsection{Four Derivative Corrections to R-Charged Black Holes}

In this section we consider a generic four derivative correction to action (\ref{gsacn}).
\ben
{\cal I} &=& {1 \over 16 \pi G_{5}}\int d^5x \sqrt{- g} \ltb R 
- {1 \over 4}X_I^2 F^I_{\mu\nu}F^{\mu\nu I}-{1 \over 2}X_I^{-2}
\partial_{\mu} X_I \partial^{\mu}X_I -V + g \,\, {\cal L}_{(4)}\rtb  \nonumber \\
&& \ \ \ \ \ \ +  C.S \,\,\, terms
\een
and
\be
 {\cal L}_{(4)}= \beta_1 R^2
+ \beta_2 R^{\mu\nu\rho\sigma}R_{\mu\nu\rho\sigma}+\beta_3 R^{\mu\nu}R_{\mu\nu}\ .
\ee
The extremal Black hole 
has $AdS_2$ part in there near horizon geometry. The full near-horizon ansatz is as follows,
\ben\label{exnh}
ds^2 &=& v_1 (- r^2 dt^2 + {dr^2 \over r^2}) + v_2 d\vec{x}^2 \nonumber \\
X_I&=& u_I, \,\,\,\ A^I= - e_I r dt,\,\,\,\ F^I_{tr}=e_I\ .
\een
We will not explain the entropy function formalism here, as it is very well studied in literature. We will just present the results here.
Following \cite{morales} we choose three parameters $\mu_1, \mu_2$ and $\mu_3$ and parametrize $u_I's$ and $v_2$ as,
\be
u_I = {\mu_I \over (\Pi \mu_J)^{1/3}}, \,\,\,\,\ v_2= (\Pi \mu_J)^{1/3}\ .
\ee
Then the other parameters are given by (in terms of $\mu_I's$),
\ben
v_1&=&{(\Pi \mu_J)^{1/3} \over 4 \gamma^2 \Sigma \mu_J}, \nonumber \\
e_I &=& e_I^0+ g e_I^g, \,\,\,\,\,\,\,\,\,\  q^I=q^I_0+ g q^I_g,\nonumber \\
e_I^0&=& {\sqrt{{\Pi \mu_J \Sigma_{J \ne I}\mu_J \over 2 \gamma^2 (\Sigma\mu_J)^2}} \over \mu_I}, \,\,\,\ 
e_I^g= - {4 (\Pi \mu_J)^{2/3} (2 \beta_1+ 2\beta_2+ \beta_3)  \over 3 \mu_I \sqrt{{\Pi \mu_J 
\Sigma_{J \ne I}\mu_J \over 2 \gamma^2 (\Sigma\mu_J)^2}}}, 
\een
and
\be\label{qi}
q^I_0 = {2 V_3 \gamma^2 \Sigma \mu_J \mu_I^2 \over (\Pi \mu_J)^{1/2}} e_I^0 ,\,\,\ 
q^I_g = {2 V_3 \gamma^2 \Sigma \mu_J
\mu_I^2 \over (\Pi \mu_J)^{1/2}} e_I^g\ .
\ee
Here, $\Sigma \mu_J= (\mu_1+\mu_2+\mu_3)$ 
and $\Pi \mu_J=\mu_1\mu_2\mu_3 $. Repeated indices in $q^I$ are $not$ summed over.

After finding higher derivative correction to the near horizon geometry we compute the entropy of the extremal Black hole,
\be
s={\sqrt{\Pi \mu_J} \over 4 G_5}\ltb 1 - 8 g \gamma^2 (2 \beta_1+2\beta_2+\beta_3)  
{\Sigma \mu_J \over (\Pi \mu_J)^{1/3}}\rtb\ + {\cal O}(g^2).
\ee

Similarly we can also find the shear-viscosity coefficient from gravitons effective coupling. 
It is given as,
\be
\eta={\sqrt{\Pi \mu_J} \over 16 \pi G_5}\ltb 1 - 8 g \gamma^2 (2 \beta_1+2\beta_2+\beta_3)  
{\Sigma \mu_J \over (\Pi \mu_J)^{1/3}}\rtb\ + {\cal O}(g^2).
\ee
In order to obtain the expressions in terms of black hole charges one has to invert the equation (\ref{qi}) and write $\mu_I's$ in terms of $q_I's$. 
The shear viscosity to entropy density ration is given by,
\be
{\eta\over s}= {1 \over 4 \pi} + {\cal O}(g^2).
\ee
Thus we see that in the extremal limit the correction saturates the {\bf KSS} bound for any generic four derivative terms in the action.

\subsection{Six Derivative Correction R-Charged Black Holes}

 In this section, we consider generic six derivative terms to the five dimensional gauge sugra Lagrangian 
and determine the $\eta/s$ ratio. There are eight 
possible six derivatives terms constructed our of curvature tensors listed below, and a generic six derivative 
correction term is constructed by taking arbitrary linear combinations of them. They are as follows,
 \ben
 T_1&=& R^{\mu\nu\sigma\kappa}R_{\sigma\kappa\rho\tau}R^{\rho\tau}_{\mu\nu}, \,\,\,\,\
 T_2= R^{\mu}_{\nu\rho\sigma}R^{\rho}_{\tau\mu\lambda}R^{\sigma\tau\nu\lambda},\nonumber \\
 T_3&=&R^{\mu\nu\sigma\kappa}R_{\sigma\kappa\nu\rho}R^{\rho}_{\mu}, \,\,\,\,\ 
T_4= R R^{\mu\nu\sigma\kappa}R_{\sigma\kappa\mu\nu}, \nonumber \\
 T_5&=&R^{\mu\nu\sigma\kappa}R_{\mu\sigma}R_{\nu\kappa}, \,\,\,\,\,\
T_6=R^{\mu}_{\nu}R^{\nu}_{\sigma}R^{\sigma}_{\mu}, \nonumber \\
 T_7&=& R R^{\mu}_{\nu}R^{\nu}_{\mu}, \,\,\,\,\,\,\ T_8= R^3, \nonumber \\
 {\cal L}_{(6)}&=& \alpha_1 T_1 +\alpha_2 T_2 +\alpha_3 T_3 +\alpha_4 T_4 
+\alpha_5 T_5 +\alpha_6 T_6 +\alpha_7 T_7 +\alpha_8 T_8
 \een

Now, we consider an action with this generic six derivative correction to (\ref{gsacn}),
\ben
{\cal I} &=& {1 \over 16 \pi G_{5}}\int d^5x \sqrt{- g} \ltb R - {1 \over 4}X_I^2 F^I_{\mu\nu}F^{\mu\nu I}-{1 \over 2}X_I^{-2}
\partial_{\mu} X_I \partial^{\mu}X_I -V + g \,\, {\cal L}_{(6)}\rtb  \nonumber \\
&& \ \ \ \ \ \ \ \ \ \ \ \ \ \ \ + C.S \,\,\, terms \ .
\een
In \cite{ns2}, we considered this correction (even with the lower order GB term) to Einstein-Hilbert action. Using 
field redefinition, we were able to show that few terms would not occur in the final expression of $\eta/s$. 
This is no longer true for this action\footnote{It is easy to check that the arguments given in reference \cite{ns2} no longer holds for all coefficients.}. The near-horizon geometry is given by (\ref{exnh}). The solution turns out to be,
\be
u_I = {\mu_I \over (\Pi\mu_J)^{1/3}}, \,\,\,\,\ v_2= (\Pi\mu_J)^{1/3}, \ \ \ v_1 = v_1^0+ g v_1^g, \,\,\,\,\,\,\,\ 
v_1^0= {v_2 \over 4 \gamma^2 \Sigma \mu_J}, \,\,\,\,\ v_1^g= -{\alpha \over v_1^0}
\ee
where,
\be
e_I=e_I^0+ g e_I^g , \,\,\,\,\,\,\,\ q^I=q^I_0+ g q^I_g\ ,
\ee
\be
e_I^0 = {\sqrt{{\Pi \mu_J \Sigma_{J \ne I}\mu_J \over 2 \gamma^2 (\Sigma\mu_J)^2}} \over \mu_I} \ , \ \ \ 
e_I^g= {8 \gamma^2 \alpha (\Pi \mu_J)^{1/3} {(\Sigma \mu_J+ 3 \mu_I)  
\over 3 \mu_I \sqrt{{\Pi \mu_J \Sigma_{J \ne I}\mu_J \over 2 \gamma^2
(\Sigma\mu_J)^2}}}}\ ,
\ee
\be
q^I_0= {V_3\mu_I^2 e_I^0 \over 2 \sqrt{v_2}v_1^0}\ , \ \ \ \ \ \ \ \ 
q^I_g= {V_3 \mu_I^2  \over 2 \sqrt{v_2}}
(e_I^g - {v_1^g \over (v_1^0)^2})
\ee
and $\alpha$ is defined to be,
\be
\alpha=4 \alpha_1 - 2 \alpha_3+4 \alpha_4 +\alpha_5+\alpha_6+2 \alpha_7+4 \alpha_8 \ .
\ee

Now, using entropy function formalism, it is easy to find out the corrected entropy (density) for the extremal 
black holes. It result is,
\be
s={\sqrt{\Pi \mu_J} \over 4 G_5}\ltb 1 + 48 g \gamma^4 \alpha {(\Sigma \mu_J)^2 \over (\Pi \mu_J)^{2/3}}\rtb + {\cal O}(g^2)\ .
\ee
 
Now, following the effective action technique, we can compute the shear-viscosity coefficient for these 
extremal black holes. The coefficient and the $\eta/s$ ratio are as follows,
\be
\eta={\sqrt{\Pi \mu_J} \over 16 \pi G_5}\ltb 1 - 16 g 
\gamma^4 (3\alpha_2+4 \alpha_3-12 \alpha_4-2 \alpha_5-3 \alpha_6-6 \alpha_7-12 \alpha_8) 
{(\Sigma \mu_J)^2 \over (\Pi \mu_J)^{2/3}}\rtb + {\cal O}(g^2)\ .
\ee
Finally $\eta/s$ is given by,
\be
{\eta\over s}={1 \over 4 \pi}\ltb 1 -16 g \gamma^4 (12\alpha_1+3 \alpha_2-2 \alpha_3+\alpha_5) 
{(\Sigma \mu_J)^2 \over (\Pi \mu_J)^{2/3}}\rtb + {\cal O}(g^2)\ .
\ee
The ratio approaches to the universal bound for pure Lovelock gravity.

\section{Black Holes in Asymptotically Lifshitz Spacetime}\label{lif}

In this section we study the hydrodynamic behaviour of boundary gauge theory plasma at Lifshitz like fixed point. The dual gravity 
theory of a Lifshitz like fixed point was first proposed by Kachru et.al. \cite{kachru}. They presented a four dimensional bulk
action in presence of one and two form fields. Although it is not possible to embed this action 
in any string theory \cite{taka} but it is a good holographic phenomenological model to study the behaviour of boundary field theory
at Lifshitz like fixed point. The action is given by,
\be
{\cal I}={1\over 16 \pi G_4}\int d4^x \sqrt{-g} \lb R - 2 \Lambda -{1\over 4} F_{\mu\nu}F^{\mu\nu}-{c^2 \over 2}{\cal A}_{\mu}{\cal A}^{\mu} + g {\cal L}_{(4)}\rb\ .
\ee
where,
\be
F=d{\cal A}
\ee
and
\be
c={\sqrt{2 z} \over L},\ \ \ \ \ \Lambda=-{z^2+z+4 \over 2 L^2} \ .
\ee
There can be order $g$ corrections to $c$ and $\Lambda$ in presence of Gauss-Bonnet term \cite{anindya}. But they will not affect $\eta/s$ calculation at order $g$. 
This action is related to \cite{kachru} up to some field re-definition.

The finite temperature solution (black hole solution) (numerically) for this kind of action was proposed in 
\cite{ulf,mann} for $z=2$ where $z$ is the critical exponent. Later in \cite{bertoldi} authors have found a numerical black hole
solution for any values of $z$. A complete analytic black hole solution for 
this kind of phenomenological action is not yet known (see \cite{taylor} for a different model). We study the near horizon 
structure of this black hole and compute entropy, shear viscosity and their ratio of the dual field theory at Lifshitz
fixed point. Our results are valid up to order $g$.

We consider the black hole solution of the following form.
\ben
ds^2 &=&-e^{2 A(r)}dt^2 + e^{2 \sigma(r)}dr^2 + r^2 (dx_1^2 + dx_2^2)\ , \nonumber \\
{\cal A}&=&e^{G(r)}dt\ .
\een
Substitute this ansatz for metric and gauge fields in the action (in $g\ra limit$) and find equations of 
motion for $A(r)$, 
$\sigma(r)$ and $G(g)$. Then we consider the following near-horizon ansatz for the fields
\ben
&&A(r)=\ln\left(r^z L\left(a_0(r-r_h)^{\frac12}+a_0a_1(r-r_h)^{\frac32}+\cdots\right)\right),\nonumber \\
&&\sigma(r)= \ln\left(\frac{L}{r}\left(c_0(r-r_h)^{-\frac12}+c_1(r-r_h)^{\frac12}+\cdots\right)\right), \nonumber 
\\ &&G(r)=\ln\left(\frac{L^2r^z}{z}\sqrt{\frac{2z(z-1)}{L^2}}\left(a_0g_0(r-r_h)+a_0g_1(r-r_h)^2+\cdots\right)\right). 
\een
and substituting them into the equations of motion we solve for different coefficients order by order and find (see \cite{bertoldi} also),
\ben \label{c0equ}
c_0 = \frac{\sqrt{(4z+g_0^2 r_h(z-1))} r_h^{\frac12}}{\sqrt{2 z}\sqrt{{{(z^2+z+4)}}}} \ ,
\een
\be
a_1=\frac{2 g_0^2 r_h \left(z^3+2 z^2+z-4\right)+g_0^4 r_h^2
   (z-1)^2-8 z^2 \left(z^2+z+4\right)}{8 r_h z
   \left(z^2+z+4\right)}\ ,
\ee
\be
c_1=\frac{-2 g_0^4 r_h^2 (z-1)^2 \left(z^2-11 z+4\right)+48 g_0^2 r_h
   (z-1) z^2+3 g_0^6 r_h^3 (z-1)^3+32 z^2 \left(z^2+z+4\right)}{8
   \sqrt{2} z^2 \left(z^2+z+4\right)^{3/2} \sqrt{\frac{r_h
   \left(g_0^2 r_h (z-1)+4 z\right)}{z}}}\ ,
\ee
\be
g_1=\frac{g_0 \left(g_0^4 r_h^2 (z-1)^2+8 g_0^2 r_h (z-1) z-4 z
   \left(z^3+2 z^2+z+4\right)\right)}{4 r_h z
   \left(z^2+z+4\right)}\ .
\ee

Having found out the leading near-horizon geometry we find entropy and shear viscosity of dual gauge theory 
at strong coupling as,
\be
\eta=\frac{r_h^2}{16 \pi  G_5}-\frac{g \left(z^2+z+4\right) \left(g_0^2 r_h (z-1) \left(2 \beta _1+2 \beta _2+\beta _3\right)+2 z \left(4 \beta _1+\beta
   _3\right)\right)}{8 \pi  G_5 L^2 \left(g_0^2 r_h (z-1)+4 z\right)}r_h^2 + {\cal O}(g^2)
\ee
and
\be
s=\frac{r_h^2}{4 G_5}-\frac{g \left(z^2+z+4\right) \left(g_0^2 r_h (z-1) \left(2 \beta _1+2 \beta _2+\beta _3\right)+2 z \left(4 \beta _1+\beta
   _3\right)\right)}{2 G_5 L^2 \left(g_0^2 r_h (z-1)+4 z\right)}r_h^2 + {\cal O}(g^2)\ .
\ee

The shear viscosity to entropy density ratio is given by,
\be
{\eta\over s}={1\over 4 \pi} + {\cal O}(g^2)+ {\cal O}(g^2)\ .
\ee
In \cite{kats} the ratio $\eta/s$ has been computed for any generic four derivative terms in any dimension for asymptotically AdS spacetime. It turns out that in four dimension the ratio saturates the universal bound. Here also we get the same result since the near horizon geometry is same for both the cases.

When we are preparing this manuscript some interesting papers appear \cite{cai3,sayan,bindu}.\\

\noindent
{\bf Acknowledgement}

We would like to that D. Astefanesei, A. Bagchi, R. Gopakumar, R. Gupta, D. Jatkar, Prem Kumar, A. Mukhopadhyaya and A. Sen for useful discussion.
This work of NB is part of the research programme of the Foundation for Fundamental Research on Matter (FOM), which is financially supported by the Netherlands Organisation for Scientific Research (NWO).
SD would like to acknowledge the hospitality of Harish-Chandra Research Institute, Allahabad, India where some part
of this work was completed.




\end{document}